\documentclass[twocolumn,twoside,aps,groupaddress,pre,showpacs,floatfix]{revtex4}

\usepackage{amsmath,amssymb,graphicx,latexsym}
\usepackage[dvips]{color}

\newcommand{\p}{\partial}

\begin{document}

\title{Suppressing the Rayleigh-Taylor instability with a rotating
  magnetic field} 

\author{Dirk Rannacher}
\email[]{rannacher@theorie.physik.uni-oldenburg.de}
\affiliation{Intitut f\"ur Physik, Universit\"at Oldenburg, 
D-26111 Oldenburg, Germany}
\author{Andreas Engel}
\affiliation{Intitut f\"ur Physik, Universit\"at Oldenburg, 
D-26111 Oldenburg, Germany}

\date{\today}

\begin{abstract}
The Rayleigh-Taylor instability of a magnetic fluid superimposed
on a non-magnetic liquid of lower density may be suppressed with the
help of a spatially homogeneous magnetic field rotating in the plane
of the undisturbed interface. Starting from the complete set of
Navier-Stokes equations for both liquids a Floquet analysis is
performed which consistently takes into account the viscosities of the
fluids. Using experimentally relevant values of the parameters we
suggest to use this stabilization mechanism to provide controlled
initial conditions for an experimental investigation of the
Rayleigh-Taylor instability.  
\end{abstract}


\maketitle

\section{Introduction}
The Rayleigh-Taylor instability \cite{Rayleigh}, \cite{Taylor},
\cite{Lewis} is a classical hydrodynamic instability
\cite{Chandrasekhar} with relevance in such diverse fields as plasma
physics, astrophysics, meteorology, geophysics, inertial confinement
fusion and granular media, for a review see, e.g.,
\cite{Sharp}. Generically this instability  
develops if a layer of liquid is superimposed to an immiscible and
less dense liquid such that the potential energy of the system can be
reduced by interchanging the liquids. Consequently, the initially plane
interface between the liquids becomes unstable and the characteristic
dimples and spikes develop resulting finally in a stable layering
with the lighter fluid on top of the heavier one.  

A quantitative experimental investigation of the Rayleigh-Taylor
instability requires reliable control of the initial
condition. Standard procedures like suddenly removed partitions
between the fluids \cite{LiRe}, \cite{Da} or quickly turning the
experimental cell upside down \cite{PlWh}, \cite{Lange} clearly
produce unpredictable initial perturbations. It is much more 
convenient to use some additional mechanism which first {\em
  stabilizes} the unstable layering of the liquids and  
may later be switched off instantaneously. It is well known that the
Rayleigh-Taylor instability may be suppressed, e.g., by vertical
oscillation of the system \cite{Wolf1}, \cite{Wolf2} and by
appropriate temperature gradients \cite{Burgess}. The first mechanism
is likely to induce uncontrolled initial surface deflections when
stopped, in the second one it is difficult to abruptly switch off the
stabilization.  

In the present paper we investigate the possibility to stabilize a
potentially Rayleigh-Taylor unstable system involving a magnetic fluid
by external magnetic fields. We will show that for experimentally
relevant parameter values moderate fields strengths which can easily
be switched on and off are sufficient to achieve the desired
stabilization. 

Magnetic fluids are suspensions of ferromagnetic nano-particles in
carrier liquids with the hydrodynamic properties of Newtonian fluids
and the magnetic properties of super-paramagnets \cite{Rosensweig},
\cite{Cowley}. It is well known that a magnetic field parallel to
the plane interface between a ferrofluid and a non-magnetic fluid
suppresses interface deflections with wave vector in the direction of
the field \cite{Rosensweig}. This may be used to stabilize the
Rayleigh-Taylor instability in two-dimensional situations where the
interface is line, as e.~g. in a Hele-Shaw cell. In the more natural 
three-dimensional setting we are interested in here a
static magnetic field parallel to the undisturbed interface is not
sufficient to stabilize the flat interface since perturbation with
wave vectors {\em perpendicular} to the magnetic field will still grow
as in the absence of the field. We therefore propose to use a spatially
homogeneous magnetic field {\em rotating} in the plane of the
undisturbed interface and determine appropriate values of the field
amplitude and rotation frequency. An alternative possibility is to use
a static {\em inhomogeneous} magnetic field with the magnetic force
counterbalancing gravity \cite{Zelazo}. This method was used in
\cite{Carles} to investigate the 2-d Rayleigh-Taylor instability in a
Hele-Shaw cell. 

Before embarking upon the detailed analysis we would like to mention
three characteristic features of our method. Firstly, we will show
that a rotating magnetic field is 
unable to suppress {\em all} possible unstable modes of the
system. In fact it can only stabilize surface deflections with 
wavenumber modulus larger than some threshold value. Perturbations
with very long wavelength are, however, not a serious problem in real
experiments because these are suppressed automatically by the finite
geometry of the sample. Secondly, it is well-known that in analogy
with the Faraday instability an oscillating magnetic field will induce
{\em new} instabilities at wave numbers which without field were stable
\cite{FaSn}, \cite{Mahr}. In our analysis we keep track of these
unstable modes and determine the magnetic field strength such that no
new instabilities may occur. For suppressing these new modes viscous
losses in the liquids will be decisive which is the reason why the
viscosities of the two liquids will be consistently taken into account
in the analysis. Finally, due to the dispersed magnetic grains
ferrofluids have usually comparatively high densities. We therefore
specialize to the case in which the upper, heavier layer is formed by the
magnetic fluid. This should be the typical situation in
experiments. Nevertheless a similar analysis with analogous results
is possible for the reverse situation with the ferrofluid at the
bottom of the system superimposed by an even denser non-magnetic liquid. 

The paper is organized as follows. In section II we collect the basic
equations and boundary conditions. In Section III we linearize these
equations around the reference state of a plane interface between the
liquids. Section IV contains the Floquet theory to determine the
boundaries separating stable from unstable regions in the parameter
plane. After shortly discussing two approximate treatments of the
fluid viscosities in section V we present the results of our analysis
in section VI. Finally section VII contains some discussion. 


\section{Basic equations}
\begin{figure}[ht]
  \includegraphics[scale=1]{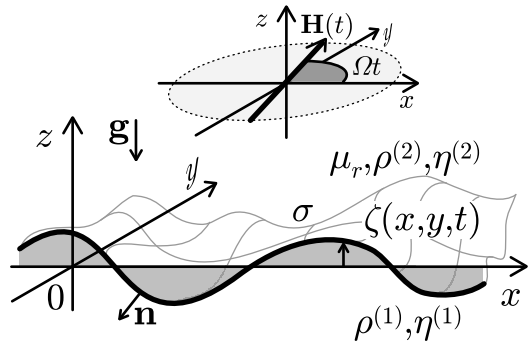}
  \caption{Sketch of the system under consideration. A ferrofluid
  of relative permeability $\mu_r$, density $\rho^{(2)}$ and viscosity
  $\eta^{(2)}$ is superimposed on a non-magnetic fluid with density
  $\rho^{(1)}<\rho^{(2)}$ and viscosity $\eta^{(1)}$. The normal
  vector ${\bf n}$ on the interface $\zeta(x,y,t)$ points into the
  non-magnetic liquid. The vector ${\bf g}$ denotes the gravitational
  acceleration, $\sigma$ is the interface tension. The system is
  subjected to a horizontal rotating magnetic field ${\bf
    H}_0=H_0(\cos(\Omega t),\sin(\Omega t), 0)$ with angular frequency
  $\Omega$.}  
  \label{rosensweig}
\end{figure}

We consider a ferrofluid with density $\rho^{(2)}$ superimposed on a
non-magnetic fluid of lower density $\rho^{(1)}<\rho^{(2)}$, see
Fig.~\ref{rosensweig}. Both layers are assumed to be infinite in
horizontal as well as in vertical direction. The densities and the
respective viscosities $\eta^{(1)}$ and $\eta^{(2)}$ are taken to be
constant. The liquids 
are immiscible and the interface between them is parametrized by
$z=\zeta(x,y,t)$. We will study the stability of a flat interface
which we take as the $x$-$y$-plane of our coordinate system, the
undisturbed interface is hence given by $\zeta(x,y,t)\equiv 0$. In the
absence of a magnetic field this situation is unstable due to the
Rayleigh-Taylor instability \cite{Rayleigh,Taylor,Chandrasekhar}. 

In the presence of an 
external magnetic field ${\bf H}$ the magnetic fluid builds up a 
magnetization ${\bf M}$ which is assumed to be a linear function of the
field, ${\bf M}=\chi{\bf H}$, where $\chi$ denotes the susceptibility
related to the relative permeability by $\mu_r=1+\chi$. Both liquids
are subject to the homogeneous gravitational field acting in negative
$z$-direction and to the interface tension $\sigma$ acting at their
common interface. The magnetic fluid is additionally influenced by the
magnetic force density $({\bf M}\nabla){\bf H}$ resulting from the
externally imposed spatially homogeneous magnetic field ${\bf H}_0 =
H_0(\cos(\Omega t), \sin(\Omega t), 0)$ rotating with constant angular
frequency $\Omega$ in the $x$-$y$-plane. 

The time evolution of the system is governed by the following set of
equations. The incompressibility of both liquids gives rise to the
continuity equations for the velocity fields ${\bf v}^{(j)}$
\begin{equation}
  \label{konti}
  \nabla\cdot{\bf v}^{(j)} = 0 \;,
\end{equation}
with $j=1,2$ where here and in the following the lower (i.e. non-magnetic)
fluid parameters are denoted with superscript $(1)$ and the higher
(magnetic) one with superscript $(2)$. The hydrodynamic equations of
motion are the Navier-Stokes equations
\begin{equation}
  \label{nse}
  \rho^{(j)} \left(\p_t + {\bf v}^{(j)}\cdot\nabla\right){\bf v}^{(j)} =
  \nabla\cdot T^{(j)} + \rho^{(j)}{\bf g}
\end{equation}
with ${\bf g}=(0,0,-g)$ denoting the acceleration due to gravity and
the stress tensors $T_{\ell m}^{(j)}$ given by 
\begin{equation}
  \label{tensor}
  \begin{split}
    T_{\ell m}^{(j)} = &-\left(p^{(j)} +
    \frac{\mu_0}{2}H^{(j)2}\right)\delta_{\ell m}\\[1ex]
    &+ B_{\ell}^{(j)}H_{m}^{(j)} + \eta^{(j)}\left(\p_\ell v_{m}^{(j)}
      + \p_mv_{\ell}^{(j)}\right) \; .
  \end{split}
\end{equation}
Here $p^{(j)}$ denotes the pressure in each liquid, and  
${\bf B}^{(j)}=\mu_0 ({\bf H}^{(j)}+{\bf M}^{(j)})$ is the respective
magnetic induction. Note that also the stress tensor for the
non-magnetic liquid contains contributions from the magnetic field
which, however, are divergence free and therefore do not give rise to
a force density in the lower fluid.  

For values of $\Omega$ relevant to the present investigation radiative
effects are negligible and the magnetic field has to obey the
magneto-static Maxwell equations
\begin{equation}
\label{maxwell}
  \begin{split}
    \nabla\cdot{\bf B}^{(j)} &= 0\\[1ex]
    \nabla\times{\bf H}^{(j)} &= 0
  \end{split}\quad.
\end{equation}
In view of the second equation it is convenient to introduce 
scalar magnetic potentials $\Phi^{(j)}$ according to 
${\bf H}^{(j)}=-\nabla\Phi^{(j)}$. These potentials then fulfill 
the Laplace equations
\begin{equation}
  \label{laplace}
  \nabla^2\Phi^{(j)} = 0 \; .
\end{equation}
The above equation have to be supplemented by appropriate boundary
conditions. Far from the interface the velocities must remain bounded,
\begin{equation}
  \label{vinfinity}
  \lim_{z\rightarrow\pm\infty}|{\bf v}^{(j)}| <\infty \; ,
\end{equation}
and the magnetic field must be equal to the externally imposed field, 
\begin{equation}
  \label{maginfinity}
  \lim_{z\rightarrow\pm\infty}-\nabla\Phi^{(j)} = {\bf H}_0 \; .
\end{equation}
To formulate the boundary conditions at the interface $z=\zeta(x,y,t)$
we define the normal vector ${\bf n}$ by
\begin{equation}
  \label{normal}
 {\bf n}=-\frac{\nabla(z-\zeta(x,z,t))}
   {\left|\nabla(z-\zeta(x,z,t))\right|} \; .
\end{equation}
The normal component of the stress tensor has to fulfill 
\begin{equation}
  \label{normaltensor}
  \sum_{\ell,m}{\mathrm n}_\ell\;\triangle T_{\ell m}\;{\mathrm n}_m =
  -\sigma K
\end{equation}
where $K=\nabla\cdot{\bf n}$ is the local curvature of the interface
and the symbol $\triangle\lambda\equiv\lambda^{(2)}-\lambda^{(1)}$
denotes here and in the following the difference in the value of the
respective quantity slightly above and slightly below the
interface. The tangential components of the stress tensor have to be
continuous, 
\begin{equation}
  \label{tangentialtensor}
  \sum_{\ell,m}{\mathrm t}_\ell\; \triangle T_{\ell m}\;{\mathrm n}_m = 0
\end{equation}
for all vectors ${\bf t}$ perpendicular to ${\bf n}$. The motion of
the interface is related to the velocity fields in the liquids by the
kinematic condition 
\begin{equation}
  \label{kinematic}
  \p_t\zeta + {\bf v}^{(j)}\cdot \nabla\zeta = {\bf v}_z^{(j)}\; .
\end{equation}
Finally, at the interface the normal component of ${\bf B}$ and the
tangential component of ${\bf H}$ have to be continuous which gives
rise to the following boundary conditions for the magnetic potentials
at the interface
\begin{equation}
  \label{magcon}
  \begin{split}
    \Phi^{(1)} &= \Phi^{(2)}\\[1ex]
    \p_n\Phi^{(1)} &= \mu_r\p_n\Phi^{(2)}
  \end{split} \quad .
\end{equation}


\section{Linear stability analysis}
The main purpose of the present work is to investigate whether the
Rayleigh-Taylor instability can be suppressed with the help of a
rotating magnetic field. We will hence study the linear stability
of the reference state with a flat interface, $\zeta(x,y,t)\equiv
0$, in dependence on the magnetic field strength $H_0$ and the angular 
frequency $\Omega$. The reference solution of the basic equations is
given by 
\begin{equation}
  \begin{split}
   & {\bf v}_0^{(j)} = 0,\quad p_0^{(j)}=-\rho^{(j)}gz,\\[1ex]
   & \Phi_0 = -H_0\left(\cos(\Omega t)x + \sin(\Omega t)y\right)
  \end{split}\qquad .
\end{equation}
To investigate its stability we introduce as usual small perturbations
\begin{equation}
  \begin{split}
    &{\bf v}^{(j)} = {\bf v}_0^{(j)} + \delta{\bf v}^{(j)},\quad p^{(j)} =
      p_0^{(j)} + \delta p^{(j)},\\[1ex]
    &\Phi^{(j)}=\Phi_0+\phi^{(j)}
  \end{split} \; ,
\end{equation}
and linearize the basic equations in these perturbations as well as in
the interface deflection $\zeta(x,y,t)$. We will denote the components
of the perturbed velocity vectors by 
$\delta{\bf v}^{(j)}=(u^{(j)},v^{(j)},w^{(j)})$.
It is convenient to introduce dimensionless quantities according to 
\begin{equation}
  \label{dimlos}
  \begin{split}
    k &\rightarrow k_{c,H=0}\;k\\[1ex]
    t &\rightarrow
    \left(\frac{\sigma}{g^3\triangle\rho}\right)^{1/4}t\\[1ex]
    p^{(j)} &\rightarrow\sqrt{\triangle\rho g \sigma}\;p^{(j)}\\[1ex]
    H^2 &\rightarrow \left(\frac{2}{\mu_0}\frac{\chi +
        2}{\chi^2}\sqrt{\triangle\rho 
      g\sigma}\right) H^2\\[1ex]
    \rho^{(j)} &\rightarrow\triangle\rho\;\rho^{(j)}\\[1ex]
    \eta^{(j)} &\rightarrow\left(
    \frac{\triangle\rho\sigma^3}{g}\right)^{1/4}\;\eta^{(j)}\; ,
  \end{split}
\end{equation}
where $k_{c,H=0}=\sqrt{\triangle\rho\,g/\sigma}$ is the critical wave
number of the Rayleigh-Taylor instability for $H_0=0$. 

The linearized set of basic equations
(\ref{konti},\ref{nse},\ref{laplace}) then reads 
\begin{align}
  \label{kontilin}
  \nabla\cdot\delta{\bf v}^{(j)} &= 0\\[1ex] 
  \label{nselin}
    \rho^{(j)}\p_t\delta{\bf v}^{(j)} &=-\nabla\delta p^{(j)}
    +\eta^{(j)}\nabla^2\delta{\bf v}^{(j)}\\ 
           &- 2\frac{\chi+2}{\chi}\nabla\left({\bf H}_0^{(j)}\cdot
             \nabla\phi^{(j)}\right) \\[1ex]  
  \label{laplacelin}
  \nabla^2\phi^{(i)} &= 0 \; .
\end{align}
In order to eliminated the pressure it is convenient to consider the
$z$-component of the curl curl of eq.(\ref{nselin}) which is of the
form 
\begin{equation}
  \label{ccnselin}
  \left(\p_t - \nu^{(j)}\nabla^2\right)\nabla^2 w^{(j)} = 0 \; ,
\end{equation}
where we have introduced the kinematic viscosities
$\nu^{(j)}=\eta^{(j)}/\rho^{(j)}$.

From the boundary conditions (\ref{vinfinity}) and
(\ref{maginfinity}) we find 
\begin{equation}
  \label{winfinity}
  \lim_{z\rightarrow\pm\infty} w^{(j)} <\infty
\end{equation} 
and 
\begin{equation}
  \label{maginfinitylin}
  \lim_{z\rightarrow\pm\infty}\p_z\phi^{(j)} = 0   \; .
\end{equation}
The boundary conditions at the interface simplify under
linearization. Generally we may replace the interface position
$z=\zeta(x,y,t)$ by $z=0$ to linear order in $\zeta(x,y,t)$. Therefore
the symbol $\triangle$ has now the more specific meaning 
$\triangle\, \lambda=\lim_{z\downarrow 0}\lambda(z)-
\lim_{z\uparrow 0}\lambda(z)$. From (\ref{kinematic}) we then get
\begin{equation}
  \label{kinematiclin}
  \p_t\zeta = w^{(j)}\Big|_{z=0}
\end{equation}
implying 
\begin{equation}
  \label{velocity1}
  \triangle\, w =0\; .
\end{equation}
Moreover, the continuity of the flow field ${\bf v}$ together with 
(\ref{kontilin}) gives rise to
\begin{equation}
  \label{velocity2}
  \triangle\; \p_z w=0 \; .
\end{equation}
From (\ref{tangentialtensor}) we find 
\begin{equation}
  \label{tangentialtensorlin}
  \triangle\Big[\eta \left(\nabla_\perp^2-\p_z^2 \right)w \Big] = 0\; .
\end{equation}
Finally, linearization of (\ref{normaltensor}) together with
(\ref{nselin}) yields 
\begin{equation}
  \label{nse+normaltensorlin}
  \begin{split}
    &\triangle\Big[\left(\rho\p_t  -\eta\left(3\nabla_\perp^2 +
      \p_z^2\right)\right)\p_z w\Big] = \\[1ex]
    &\quad\nabla_\perp^2\left(\left(1 + \nabla_\perp^2\right)\zeta +
    2\frac{\chi+2}{\chi}{\bf H}_0\cdot\nabla\phi^{(2)}\Big|_{z=0}\right) \; .
  \end{split}
\end{equation}
where the horizontal Laplace operator is defined by $\nabla_\perp^2
=\p_x^2+\p_y^2$. 

The magnetic boundary conditions (\ref{magcon}) acquire the form
\begin{equation}
  \label{magconlin}
  \begin{split}
    &\triangle\, \phi=0\\[1ex]
    &M_0\Big(\cos(\Omega t)\p_x\zeta + \sin(\Omega t)\p_y\zeta\Big) +
    \p_z\Big(\mu_r\phi^{(2)}-\phi^{(1)}\Big)=0\; .
  \end{split}
\end{equation}

To find a solution of the set of linearized equations
(\ref{laplacelin}), (\ref{ccnselin}) together with their boundary
conditions we may exploit their translational invariance and have to
keep in mind their explicit time dependence induced by the second
boundary condition (\ref{magconlin}) for the magnetic field
problem. An appropriate ansatz is therefore given by 
\begin{equation}
  \label{ansatzes1}
  \left(
  \begin{array}{c}   
    \zeta(x,y,t)\\[1ex] 
    w^{(j)}(x,y,z,t)
  \end{array}\right)=
  \left(
  \begin{array}{c}
    \hat\zeta(t)\\[1ex]
    \hat w^{(j)}(z,t)
  \end{array}\right)e^{i(k_xx + k_yy)}
\end{equation}
and 
\begin{align}
  \label{ansatzes2}
  \phi^{(1)}(x,y,z,t)&=\hat\phi(t)\; e^{i(k_xx + k_yy) + kz}\\
   \label{ansatzes3}
  \phi^{(2)}(x,y,z,t)&=\hat\phi(t)\; e^{i(k_xx + k_yy) - kz}\; .
\end{align}
With the abbreviation $k^2=k_x^2 + k_y^2\;$ eq.~(\ref{ccnselin}) 
acquires the form  
\begin{equation}
\label{nselinf}
  \Big(\p_t -
    \nu^{(j)}\left(\p_z^2-k^2\right)\Big)\Big(\p_z^2-k^2\Big)\hat
  w^{(j)}(z,t) = 0 \; .
\end{equation}
Moreover the ansatzes (\ref{ansatzes2}) and (\ref{ansatzes3}) already
fulfill (\ref{laplacelin}), (\ref{maginfinitylin}) and the first of
the boundary conditions (\ref{magconlin}). The second one yields  
\begin{equation}
  \label{phi}
  \hat\phi(t) = i\frac{\chi}{\chi +
    2}H_0\frac{\hat\zeta(t)}{k}\left(\cos(\Omega t)k_x +
  \sin(\Omega t)k_y\right)\, ,
\end{equation}
which gives rise to 
\begin{equation}\label{h1}
  {\bf H}_0\cdot\nabla\phi^{(2)}\Big|_{z=0}=-\frac{\chi}{\chi+2}H_0^2
  k\cos^2(\Omega t)\hat\zeta(t)e^{i(k_xx+k_yy)}\,.
\end{equation}
The boundary conditions 
(\ref{winfinity},\ref{kinematiclin}-\ref{nse+normaltensorlin}) 
assume the form
\begin{align}
\label{winfinity2}
  \lim_{z\rightarrow\pm\infty} \hat w^{(j)}& <\infty\\[1ex]
\label{kinematiclinf}
  \p_t\hat \zeta &= \hat w^{(j)}\Big|_{z=0}\\
  \label{velocity1f}
  \triangle\, \hat w &= 0\\[1ex]
  \label{velocity2f}
  \triangle\, \p_z \hat w &= 0\\[1ex]
\label{tangentialtensorlinf}
 \triangle\Big[\eta \left(k^2 + \p_z^2\right)\hat w\Big] &= 0
\end{align}
and, using also (\ref{h1}), 
\begin{equation}
  \label{nse+normaltensorlinf}
  \begin{split}
    \triangle\Big[&\left(\rho\p_t -
        \eta\left(\p_z^2-3k^2\right)\right)\p_z\hat
        w\Big]=\\[1ex]
          &\Big(-1 + H_0^2(1+\cos(2\Omega t))k +
          k^2\Big)k^2\hat\zeta\; .
  \end{split}
\end{equation}
We now invoke Floquet theory \cite{JordanSmith,KumarTuckerman} to
solve this system of linear differential equations with time periodic
boundary conditions for the amplitudes $\hat w^{(j)}(z,t)$ and $\hat
\zeta(t)$.  


\section{Floquet theory}
In order to analyze the stability of the flat interface we employ the
following Floquet ansatz for the time dependence of the interface perturbation
amplitude $\hat\zeta$ and the $z$-components of the velocity $\hat w^{(j)}$:
\begin{equation}
  \label{floquetansatz}
  \Big\{\hat \zeta(t),\hat w^{(j)}(z,t)\Big\}
  =e^{(\alpha+i\beta)\,\Omega t}\sum_n\Big\{\tilde\zeta_n,\tilde
  w_n^{(j)}(z)\Big\} e^{2in\,\Omega t} \, ,
\end{equation}
where $\alpha + i\beta$ is the Floquet exponent. Here $\alpha$ is a
real number and negative $\alpha$ describes stable situation whereas
positive $\alpha$ signals an instability of the reference state. The
imaginary part $\beta$ of the Floquet exponent is either zero or one
and distinguishes between harmonic ($\beta=0$) and subharmonic
($\beta=1$) response of the system \cite{JordanSmith}. Plugging
(\ref{floquetansatz}) 
into (\ref{nselinf}) we find 
\begin{equation}
  \label{dglw}
  \big(\p_z^2 - q_{n}^{(j)2}\big)\big(\p_z^2-k^2\big)\tilde
  w_{n}^{(j)}(z) = 0 \; ,
\end{equation}
where
\begin{equation}
  q_n^{(j)}=\sqrt{k^2+\frac{\alpha + i(\beta+2n)}{\nu^{(j)}}\Omega}\;.
\end{equation}
Eq.(\ref{dglw}) has the solution 
\begin{equation}
  \tilde w_{n}^{(j)}(z) = A^{(j)}_n\,e^{kz} + B^{(j)}_n\,e^{-kz} + 
       D^{(j)}_n\,e^{q_{n}^{(j)}z} + C^{(j)}_n\,e^{-q_{n}^{(j)}z}\, ,
\end{equation}
where the constants $A^{(j)}_n ... D^{(j)}_n$ can be determined with
the help of the boundary conditions
(\ref{winfinity2}-\ref{tangentialtensorlinf}). As a result the
amplitude of the $z$-component of the velocity may be expressed in
terms of the interface amplitude $\zeta_n$ according to  
\begin{equation}
  \label{velocitysol}
  \begin{split}
    \tilde w_n^{(2)} &= \left(\tilde B_ne^{-kz} + \tilde
      D_ne^{-q_n^{(2)}z}\right)\tilde\zeta_n\\[1ex] 
    \tilde w_n^{(1)} &= \left(\tilde A_ne^{kz} + \tilde
      C_ne^{q_n^{(1)}z}\right)\tilde\zeta_n \; ,
  \end{split}
\end{equation}
where
\begin{widetext}
  \begin{equation}
    \begin{split}
      \tilde A_n &= \frac{-\triangle\eta k^4
        - \left(\eta^{(1)}q_n^{(1)} +
        \eta^{(2)}q_n^{(2)}\right)q_n^{(1)}q_n^{(2)2} -
      \eta^{(1)}\triangle q_n^2 k^2 +
        \eta^{(2)}\left(q_n^{(1)}+q_n^{(2)}\right)
        \left(k^3+\left(k-q_n^{(2)}\right)q_n^{(2)}k\right)}  
      {\left(k-q_n^{(1)}\right)\left(\eta^{(2)}\left(k+q_n^{(2)}\right) +
        \eta^{(1)}\left(k+q_n^{(1)}\right)\right)}\nu^{(2)}\\[1ex]
      \tilde B_n &= \frac{\triangle\eta k^3 + \left(\eta^{(1)}q_n^{(1)}
        +\eta^{(2)}q_n^{(2)}\right)k^2 +
      \left(\eta^{(1)}\left(k+q_n^{(1)}\right) + 
        \eta^{(2)}\left(k+q_n^{(2)}\right)\right)q_n^{(2)2} +
        2\eta^{(1)}q_n^{(1)}q_n^{(2)}k}{\eta^{(2)}\left(k + q_n^{(2)}\right) +
        \eta^{(1)}\left(k + q_n^{(1)}\right)}\nu^{(2)}\\[1ex]
      \tilde C_n &=\frac{\left(\eta^{(1)}k
          +\eta^{(2)}q_n^{(2)}\right)\left(q_n^{(2)2} - k^2\right)} 
      {\left(k-q_n^{(1)}\right)\left(\eta^{(2)}\left(k+q_n^{(2)}\right) +
        \eta^{(1)}\left(k+q_n^{(1)}\right)\right)} 2k\nu^{(2)}\\[1ex]
      \tilde D_n &= -\frac{\left(\eta^{(1)}q_n^{(1)}
          +\eta^{(2)}k\right)\left(k + q_n^{(2)}\right)} 
      {\eta^{(2)}\left(k + q_n^{(2)}\right) + \eta^{(1)}\left(k +
          q_n^{(1)}\right)}2k\nu^{(2)} \quad .
    \end{split}  
  \end{equation}
Finally, using these results in (\ref{nse+normaltensorlinf}) we find a
relation of the form
\begin{equation}
  \label{finaleq}
    \sum_{n=-\infty}^{\infty} \left\{W_n\tilde\zeta_n -
    H_0^2\left[\tilde\zeta_n +\frac{1}{2}(\tilde\zeta_{n-1} +
      \tilde\zeta_{n+1})\right]\right\} k^3\, e^{(\alpha+i(\beta
    +2n))\,\Omega t}=0 \; .
\end{equation}
Since this equation has to hold for all values of $t$ all coefficients
in the curly brackets must vanish separately. We therefore end up with
an infinite homogeneous system of linear equations for the amplitudes
$\tilde\zeta_n$ in which the off-diagonal terms arise due to the time
dependence in (\ref{nse+normaltensorlinf}). Nontrivial solutions for
the $\tilde\zeta_n$ require that the determinant of the coefficient
matrix vanishes which yields the desired relation between the
parameters of the problem, $H_0$, $k$ and $\alpha$. In the present
investigation we are mainly interested in the stability boundaries in
the parameter plane. We therefore specialize to the case $\alpha=0$
and find for the coefficients $W_n$ in (\ref{finaleq})
  \begin{equation}
    \label{wn}
    \begin{split}
      W_n = &\frac{1}{k^3}\left[-i(2n +
          \beta)\,\Omega\left(\rho^{(2)}(k\tilde B + q_n^{(2)}\tilde D) +
          \rho^{(1)}(k\tilde A + q_n^{(1)}\tilde C)\right) +
          \eta^{(2)}(k^3\tilde B + q_n^{(2)3}\tilde D) + \eta^{(1)}(k^3\tilde A
          + q_n^{(1)3}\tilde C)\right]+\\[1ex] & \frac{1}{k}\Big[-
          3\left(\eta^{(2)}(k\tilde B +
          q_n^{(2)}\tilde D) + \eta^{(1)}(k\tilde A + q_n^{(1)}\tilde
          C)\right) + 1 - k^2\Big]\; .
    \end{split}
  \end{equation}
\end{widetext}
To exploit the solvability condition for a numerical determination of
the stability boundaries we have to truncate the infinite system of
linear equations at some finite value $n_{\mathrm{max}}$ of
$n$. Comparing the results for different values of $n_{\mathrm{max}}$
the accuracy of the procedure may be estimated. For the results
presented in section VI we have used $n_{\mathrm{max}}=19$, i.e. we
have included 39 terms, $-19\leq n \leq 19$. 


\section{Special cases}
Before presenting explicit results of our analysis for experimentally
relevant parameter values it is instructive to consider two limiting
cases for which alternative approaches are available. Let us first
discuss the situation of ideal liquids,
$\eta^{(1)}=\eta^{(2)}=0$. Using (\ref{nselinf}), (\ref{kinematiclinf}) 
and (\ref{velocity1f}) we may then express $\hat w^{(j)}$ in terms
of $\hat\zeta$. Plugging the result into the boundary condition 
(\ref{nse+normaltensorlinf}) we obtain the following Mathieu equation
for the amplitude of the surface deflection $\hat\zeta(t)$:
\begin{equation}
  \label{mathieu}
  \begin{split}
    \p_t^2\hat\zeta +  \frac{\hat\zeta}{\rho_1 +
      \rho_2} &\Big(-k + H_0^2 k^2 + k^3\\[1ex] 
    &+ H_0^2 k^2\cos(2\Omega t)\Big)=0\; .
  \end{split}
\end{equation}
From the standard stability chart of the Mathieu equation \cite{AbSt}
we are now able to determine the threshold for the amplitude $H_0$ of
the external field necessary to stabilize interface deflections with
wavenumber modulus $k$. However, since most ferrofluids are rather
viscous this theory will not adequately describe the experimental
situation. 

It is possible to approximately incorporate the influence of viscosity
by assuming that the dominant contribution to viscous
damping originates far from the interface in the bulk of the fluids
where the flow field is identical to the one of ideal liquids 
\cite{Lamb,LL}. One may then derive a damped Mathieu equation for the
amplitudes of the interface deflection of the form
\begin{equation}
  \label{mathieudamped}
  \begin{split}
    \p_t^2\hat\zeta + 2\gamma\p_t\hat\zeta + \frac{\hat\zeta}{\rho_1 +
      \rho_2} &\Big(-k + H_0^2 k^2 + k^3\\[1ex] 
    &+ H_0^2 k^2\cos(2\Omega t)\Big)=0 \; ,
  \end{split}
\end{equation}
where the damping constant $\gamma$ is given by 
\begin{equation}
  \gamma= 2\frac{\eta_1 + \eta_2}{\rho_1 + \rho_2}k^2 \; .
\end{equation}
Since the damped Mathieu equation may be mapped on the undamped one
\cite{AbSt} we may again employ the stability chart of the Mathieu
equation to discuss the stabilization of a surface deflection mode
with wave vector modulus $k$ in the presence of small damping. In the
following section we compare the results of these approximate
estimates with those of our complete treatment.


\section{Results}
\begin{figure}[ht]
  \includegraphics[scale=0.6]{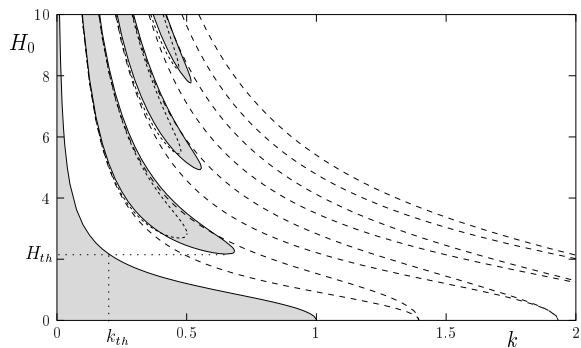}
  \caption{\label{Hk-1} Stability chart in the plane spanned by the
    dimensionless wavenumber $k$ and the magnetic field amplitude
    $H_0$ for the flat interface between a ferrofluid and a
    non-magnetic fluid in a rotating magnetic field. The dimensionless
    angular frequency of the magnetic field is $\Omega=0.69$
    corresponding to $f=10\,{\rm Hz}$, the remaining parameters are as
    given in the  main text. The full lines separating the white
    regions of stability from the gray regions of unstable
    combinations derive from a numerical solution of (\ref{finaleq}).
    For comparison also the results of the inviscid theory 
    building on (\ref{mathieu}) and of the approximate treatment of
    viscosity related to (\ref{mathieudamped}) are included as long
    and short dashed lines respectively. 
    The threshold values of the magnetic field and wave number are
    $H_{th}=2.2$ and $k_{th}=0.2$ corresponding in physical units to
    $H_{th}\simeq 7.3\,{\rm kAm^{-1}}$ and $\lambda_{th}\simeq 3.7\,{\rm cm}$.}
\end{figure}
In this section we display detailed results of our analysis for a
typical experimental combination of a ferrofluid and an immiscible
non-magnetic fluid which has been used in a related experimental
investigation \cite{Pacitto}. The fluid parameters are as follows: 
$\rho^{(2)}=1690\,{\rm kg\, m^{-3}}$, $\eta^{(2)}=0.14\,{\rm Pa\, s}$,
$\chi=2.6$, $\rho^{(1)}=800\,{\rm kg\, m^{-3}}$,
$\eta^{(1)}=0.005\,{\rm Pa\, s}$, and $\sigma=0.012\,{\rm
  N\,m^{-1}}$. For the capillary length $\lambda_c=2\pi/k_{c,H=0}$ we
then obtain $\lambda_c\simeq 7\,{\rm mm}$. The dimensionless magnetic
field amplitude $H_0=1$ corresponds to a field of $H_0=3.3\,{\rm kA/m}$,
$\Omega=1$ corresponds to a field rotating with frequency
$f=14.6\,{\rm Hz}$. 

\begin{figure}[ht]
  \includegraphics[scale=0.6]{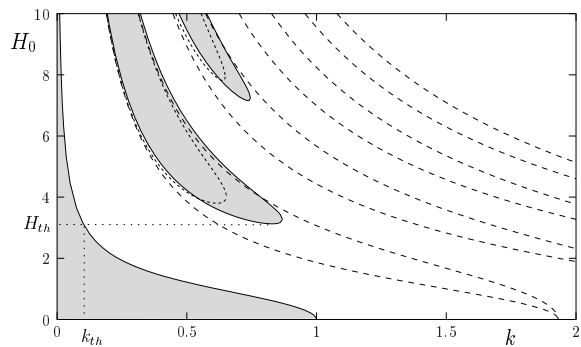}
  \caption{\label{Hk-2} Same as Fig.~\ref{Hk-1} for an 
  angular frequency of $\Omega=1.37$ ($f=20\,{\rm  Hz}$). Due to the
  higher frequency the influence of the viscosity is stronger giving
  rise to the threshold values $H_{th}=3.1$ and $k_{th}=0.1$
  corresponding to  $H_{th}\simeq 10.3\,{\rm kAm^{-1}}$  and $\lambda_{th}\simeq
  7.4\,{\rm cm}$.}
\end{figure}

In Figs.~\ref{Hk-1} and \ref{Hk-2} we show the regions of instability
of the flat interface in the $k$-$H_0$ plane. For $H_0=0$ all
perturbations are unstable for which the modulus of the wave vector is
smaller than 1 (in our dimensionless units, cf. (\ref{dimlos})), which
is the well-known trademark of the Rayleigh-Taylor
instability. Increasing $H_0$ from zero the interval of unstable wave
numbers shrinks and hence more and more long-wave perturbations may be
stabilized. However, if $H_0$ gets larger than a threshold value
$H_{th}$ the parametric excitation due to the time-dependent magnetic
field gives rise to new instabilities at higher wave-numbers. Since
these additional unstable modes are clearly unwanted $H_0$ must remain
below this threshold value $H_{th}$. Correspondingly there is a
threshold $k_{th}$ for the wavenumber modulus such that perturbations
with $k< k_{th}$ cannot be stabilized with the help of the magnetic
field. As we will detail in section VII these modes have to be
stabilized by lateral boundary conditions. We note that with
decreasing $\Omega$ the tongues of instability move closer together
and come nearer to the $k$-axis implying $H_{th}\to 0$ and $k_{th}\to
1$ for $\Omega\to 0$. 

It is clearly seen from the figures that the stability regions are
strongly influenced by the viscosity of the liquids. In the inviscid
theory the tongues of instability all reach the $k$-axis implying that
any rotating magnetic field would induce new unstable modes at values
of $k$ that were stable in the absence of the field. Therefore a
complete suppression of the Rayleigh-Taylor instability would be
impossible. It is also apparent that for realistic parameter
combinations the phenomenological inclusion of
viscosity in the theoretical description as discussed in the previous
section may give results which significantly differ from the complete
theory. This is similar to the analysis of the Faraday instability
performed in \cite{KumarTuckerman}. 

\begin{figure}[ht]
  \includegraphics[scale=0.6]{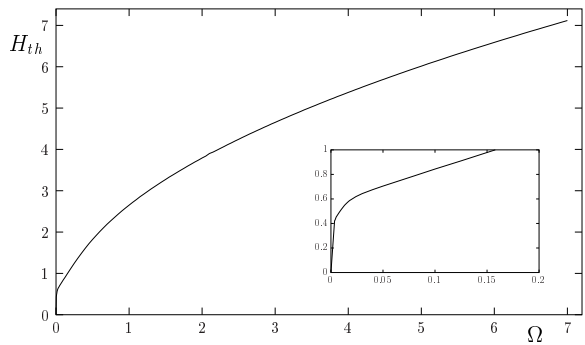}
  \caption{\label{OHc} Threshold value $H_{th}$ of the magnetic field
    amplitude $H_0$ as function of the angular frequency $\Omega$ of
    the field for the parameters given in the main text. The displayed
    interval $\Omega=1...7$ corresponds in physical units to 
    $f\simeq 15\,{\rm Hz}... 100\,{\rm Hz}$. The inset shows a blow-up
    of the steep increase of $H_{th}$ for small values of $\Omega$.}
\end{figure}

\begin{figure}[ht]
  \includegraphics[scale=0.6]{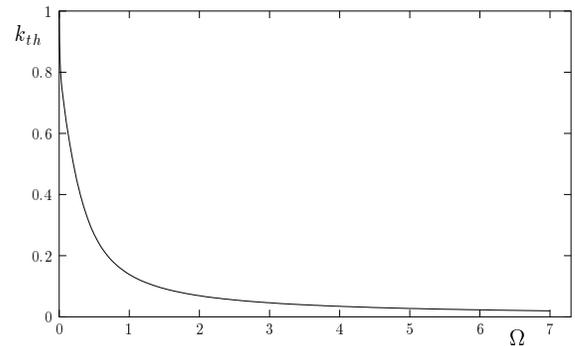}
  \caption{\label{omegakc} Similarly to Fig.~\ref{OHc} the threshold
    value $k_{th}$ of the wavenumber modulus $k$ is shown as function
    of the angular frequency $\Omega$ of the field.}
\end{figure}

Figs.~\ref{OHc} and \ref{omegakc} display the dependence of the
threshold values $H_{th}$ and $k_{th}$ on the angular frequency
$\Omega$ of the field. Clearly $H_{th}$ increases and $k_{th}$
decreases with increasing $\Omega$ as exemplified also by a comparison
between Fig.~\ref{Hk-1} and \ref{Hk-2}. For the parameters considered
an increase in $\Omega$ beyond $\Omega=2$ does not significantly
reduce $k_{th}$ any more. 

\begin{figure}[ht]
  \includegraphics[scale=0.6]{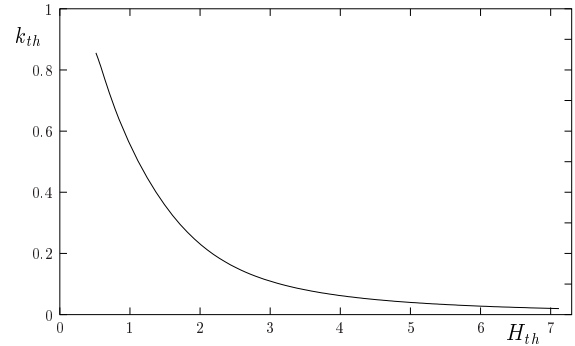}
  \caption{\label{Hckc} Threshold value $k_{th}$ of the wave number
    $k$ versus threshold value $H_{th}$ of the magnetic field
    amplitude $H_0$, again for the special set of parameters given in
    the main text. }
\end{figure}

Finally, Fig.~\ref{Hckc} combines Figs.~\ref{OHc} and \ref{omegakc}
and shows the relation between the two threshold values $H_{th}$ and
$k_{th}$. 


\section{Discussion}
In the present paper we have investigated the possibility to
stabilize a layering of a ferrofluid and a non-magnetic fluid which
were potentially unstable due to the Rayleigh-Taylor instability by a
spatially homogeneous magnetic field rotating in the plane of the
undisturbed interface. Special emphasis was put on an exact treatment of the
influence of the viscosities by starting from the complete set of
Navier-Stokes equations for both liquids. Our results show that this 
approach is for experimentally relevant parameter values superior to
both the inviscid theory and to a standard phenomenological procedure
to include viscous effects using the inviscid flow field. 

The trademark of the Rayleigh-Taylor instability is a band of unstable
wave numbers extending from $k=0$ up to a threshold value $k_{th}$
which in the absence of magnetic effects is given by the capillary
wavelength $k_c=\sqrt{\triangle \rho g/\sigma}$. The main result of
the present investigation is that $k_{th}$ may be reduced roughly by a
factor of ten with the help of a rotating magnetic field of
experimentally easily accessible amplitude and frequency. As expected
the stabilization works best for ferrofluids with high susceptibility
$\chi$ which, however, have also high densities increasing in turn 
$k_c$. 

In order to provide a clean initial condition for an experimental
study of the onset of the Rayleigh-Taylor instability one has also to
stabilize the modes with $k\leq k_{th}$. One way to
accomplish this suppression may be to use the boundary condition of a
finite geometry, i.e. by pinning the contact line between the liquids
at the boundary of the sample. In this way all long wave-number modes
with $k<k_{bc}$ are stabilized. Here $k_{bc}$ is determined by the
linear extension $L$ of the sample and roughly given by $k_{bc}\simeq
\pi/L$. Modes with $k>k_{c,H=0}$ are suppressed by surface tension. If one
is able to temporarily stabilize the remaining modes by the rotating
magnetic field, i.e. if one is able to realize $k_{th}<k_{bc}$ the flat
interface is stable.  Switching off the magnetic field at a given time
all modes with $k_{bc}<k<1$ will become unstable. Since it is
easily possible to realize values of $k_{th}$ significantly smaller than
$k_{max}$, the wavenumber with largest growth rate in the absence of
the field, the ensuing Rayleigh-Taylor instability should closely
resemble the case without lateral boundary conditions. 
To be precise it should be emphasized that our theoretical
analysis is for infinite layers only and does not take into account
the influence of lateral boundary conditions. However, the relevant
values of $H_{th}$ and $k_{th}$ will only marginally be modified.  

For an order of magnitude estimate let us consider a cylindrical
vessel of diameter $d=5\,{\rm cm}$. Pinning the contact line at the
boundary the instability of modes with dimensionless wavenumber 
$k<k_{bc}\simeq 0.11$ will be suppressed. On the other hand a rotating
magnetic field with amplitude $H_0 \simeq 10\,{\rm kAm^{-1}}$ and
frequency $f=20\,{\rm Hz}$ realizes $k_{th}\simeq 0.1$
(cf. Fig.~\ref{Hk-2}). Switching off the magnetic field all
modes with $k_{bc}<k<1$ will become unstable. For the above example
this includes the first 8 cylindrical modes which should allow a
rather accurate study of the Rayleigh-Taylor instability. We hope that
our theoretical study may stimulate experimental work along these lines. 

\begin{acknowledgments}
We would like to thank Konstantin Morozov for helpful discussion.
\end{acknowledgments}


\end{document}